# The interesting case of a single-junction solar cell in outer space

Ido Frenkel & Avi Niv*

Department of Solar Energy and Environmental Physics, Jacob Blaustein Institutes for Desert Research, Ben-Gurion University of the Negev, Sede Boqer Campus, 8499000, Israel

* Corresponding author: aviniv@bgu.ac.il

## Abstract

An isolated single-junction solar cell's temperature in outer space depends only on its radiation exchange with its environment. We consider a cell with zero emissivity below its bandgap and unity above it – an idealization so far considered to define the upper limit of photovoltaic power conversion efficiency. For this case, we show that the detailed-balance and the energy-conservation laws that govern the cell's state—temperature and potential, cannot be mutually solved. However, the cell's temperature and potential can be determined if a finite amount of sub-bandgap emissivity is included, the exact amount of which is found by minimizing the process's entropy generation. Finally, we generalize this result to a photovoltaic system in contact with some environment, hence for terrestrial conditions. Therefore, a universal thermodynamic formulation of the photovoltaic effect emerges. Unlike former attempts to thermodynamically justify the PV effect, our formalism applies to a work producing system.

## Introduction

In their seminal paper[1], Shockley and Queisser (SQ) established the limits of power conversion efficiency of a single junction (photovoltaic) solar cell based on the principle of detailed balance (DB)—the equality of electron-hole pairs generation and recombination rates. There, only the simplest case of a step emissivity, which is zero below the bandgap and unity above it, was considered – an emissivity function that defines the highest potential or work-production from a PV process[2,3]. Therefore, the step emissivity is regarded as the ultimately allowed limit – more realistic conditions would only degrade the PV process.

The DB photovoltaics law for the step emissivity is as follows[1]:

$$\Omega_s N_g^\infty(0, T_s) = \Omega_c N_g^\infty(V, T) + I. \tag{1}$$

Here $\Omega_s$, $\Omega_c$, are the solid angles of sunlight- and cell-radiation, respectively. On the left is the generation rate $N_g^\infty(0, T_s)$ of electron-hole pairs due to the perfect above bandgap absorption of blackbody radiation at a temperature $T_s$ (sunlight, which has zero potential), and $g$ is the material energy bandgap. On the right are the radiative recombination rate $N_g^\infty(V, T)$ of electron-hole pairs at a potential $V$ (in energy units eV) and temperature $T$, and electrical current $I$ (current henceforth). The radiative generation or recombination rates are given by the generalized Plank formula [4]:

$$N_a^b(V, T) \equiv \frac{2e}{c^2 h^3} \int_a^b \frac{E^2 dE}{exp\left(\frac{E-V}{kT}\right) - 1}. \tag{2}$$

Here $c$, $h$, and $k$ as the light-speed, Plank's and Boltzmann's constants, respectively, $e$ is the fundamental charge, and $a$ and $b$ are arbitrary integration limits ($g$ to $\infty$ in Eq. (1)). The SQ approach finds the cell's potential by solving Eq. (1) for a given temperature, which is often taken to be that of the cell's immediate environment. In this case, the temperature is not a free-variable of the PV process but rather an external parameter imposed on the system.

In the following, we consider a solar cell in outer space, namely an isolated cell without material contact to an environment. In this case, the cell's temperature cannot be pre-determined, as SQ does, but emerges from its radiative heat exchange with a source – the sun in this case. We show that the radiative exchange and the DB laws that govern the temperature and potential of the PV process cannot be mutually solved for a step emissivity. This inconsistency raises the possibility that this form of emissivity is perhaps an oversimplification. To find if this is indeed the case, we relax this stringent condition somewhat by allowing some sub-bandgap emissivity (SBG) and show that a solution does emerge in this case. Notice that SBG has been observed in semiconductors at elevated temperatures[5–7], but its thermodynamic role has so far remained obscure. The consideration of an arbitrary amount of SBG raises the question of its allowed value, which we answer by minimizing the PV process's entropy generation. As a result, a unification of thermodynamics with the DB law emerges. This unification goes beyond previous attempts that considered only open-circuit conditions ($I = 0$) by allowing also work producing systems[8–12]. Finally, we generalize our result to include heat conduction or convection to some environment; hence a universal thermodynamic treatment of the PV process emerges. We then demonstrate our approach with a few examples that help us identify some key aspects of our formalism, such as when it reproduces the SQ results. Our unified theory is particularly relevant for high-temperature PVs schemes such as thermophotovoltaic and thermophotonic approaches[13,14], or microcell concentrated photovoltaics for space missions[15]

## The shortcoming of the existing approach

Let us examine the case-study of an isolated solar cell in outer space. In this case, the cell has no heat-transferring contact with a thermostat (heat-bath) like the terrestrial environment. Let us further assume the cell is facing the sun and that it has no blackbody emission from its back (its back facet can be painted white, for example). Under these circumstances, let us ask ourselves, what is the open-circuit potential ($V_{OC}$ at $I = 0$) that this cell will develop by being exposed to the sun? It is clear by now that the potential cannot be solely determined from the DB law of Eq. (1) since the cell's temperature is unknown. The lack of a thermostat, in this case, has promoted the cell's temperature $T$ from being an external parameter to become the second free-variable of the photovoltaic process, in addition to the existing one, which is $V$.

To find the extraterrestrial cell's temperature, we need to know the radiative heat exchange with its environment—the energy balance (EB) or thermodynamics' first law. For the extraterrestrial cell at open circuit ($I = 0$), this would be:

$$\Omega_s E_g^\infty(0, T_s) = \Omega_c E_g^\infty(V, T). \tag{3}$$

On the left, we have the radiative heat flux absorbed by the cell from the sun, and on the right is the cell's emission. The energy flux integrals are given by:

$$E_a^b(V, T) \equiv \frac{2e}{c^2 h^3} \int_a^b \frac{E^3 dE}{exp\left(\frac{E-V}{kT}\right) - 1}. \tag{4}$$

Seemingly, we now have two equations, (1) and (3), to determine the two unknown free variables of the cell, $T$ and $V$, which is, after all, a common approach for elevated temperature PV analysis and similar systems[16–18]. Unfortunately, for our case-study, Eqs. (1) and (3) cannot be mutually solved over thermodynamic-acceptable values of $T$ and $V$. To show this insolvability, we plot in

Figure 1 $V(T)$ of the DB law from Eq. (1) in blue and that from the EB law of Eq. (3) in red. The system parameters in this case mimic a silicon cell at an earth-like orbit around the sun, such that $\Omega_s = 6.87 \times 10^{-5}$ sr, $\Omega_c = \pi$, $g = 1.12$ eV, and $T_s = 5778$ K. It is straightforward to see that there is no $V(T)$ that mutually solves Eqs. (1) and (3), the governing laws of the PV effect, for $0 < T < T_s$ and $0 < V < E_g$. One, therefore, cannot determine the temperature $T$ or the potential $V$ from Eqs. (1) and (3) for the idealized case of an isolated cell with a step emissivity.

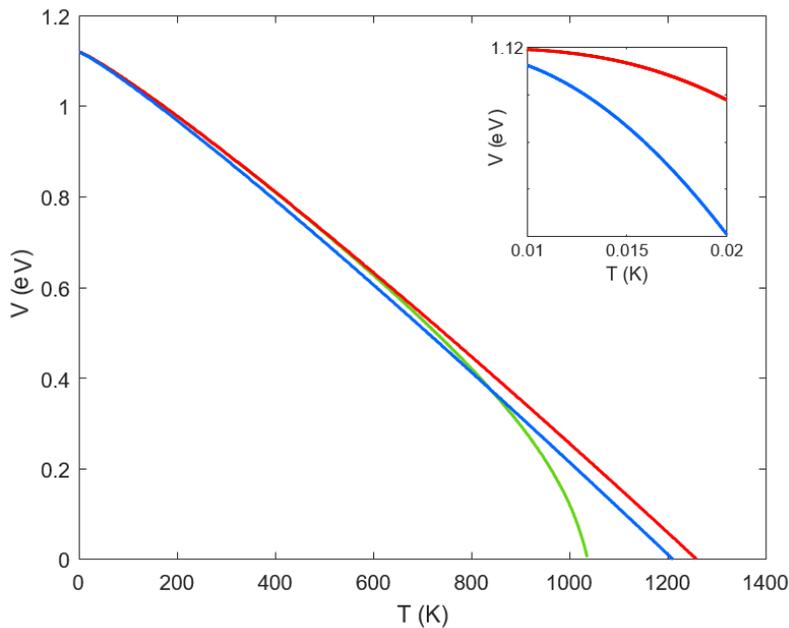

**Figure 1:** The potential as a function of the temperature from the DB relation from Eq. (1) (blue curve) and the EB relation from Eq. (3) (red curve). Insets show that the blue (DB) and red (EB) curves do not intercept even at low temperatures. The green curve shows the EB solution with $\varepsilon_{SBG} = 0.015$ in Eq. (5).

## The necessity of sub-bandgap emissivity

Faced with the inability to solve the DB and EB laws, we raise the possibility that perhaps the stepwise emissivity, namely a perfect absorption above the bandgap and none below it, is an oversimplification, at least from a thermodynamic perspective. Thus, a finite amount of sub-bandgap (SBG) emissivity is, perhaps, necessary. If successful, the consideration of SBG would establish a new idealized operational thermodynamic limit for the PV process. The microscopic origin of the SBG may be any intra-band dissipation mechanism such as carrier-carrier scattering[5,19]. According to the fluctuation-dissipation theorem[20,21], dissipation inevitably causes charge-fluctuations that act as sub-bandgap radiation sources in this case[22]. However, from the thermodynamic perspective, the microscopic origin of SBG is not essential and may vary from one PV system to another; what matters is only its inclusion dissolves the DB-EB inconsistency.

Let us reformulate the EB law to include the added energy exchange due to $\varepsilon_{SBG}$ — the hypothesized SBG:

$$\Omega_s E_g^\infty(0, T_s) + \varepsilon_{SBG}\Omega_s E_0^g(0, T_s) = \Omega_c E_g^\infty(V, T) + \varepsilon_{SBG}\Omega_c E_0^g(0, T), \qquad (5)$$

The green line in Figure 1 shows the solution of Eq. (5) for $\varepsilon_{SBG} = 0.015$, while all other parameters are kept the same. It is seen that now an intersection emerges with the DB solution of Eq. (1) at a temperature a little over 800 K and $V = 0.4$ eV. It is important to note that $\varepsilon_{SBG}$ does not affect the DB law of Eq. (1) that only concerns processes from above the bandgap energy. Also, the choice of $\varepsilon_{SBG} = 0.015$ is arbitrary and only brought as an example for a solution of the two governing laws of the PV effect. In fact, any $0 < \varepsilon_{SBG} < 1$ could be inserted.

### The irreversible generated entropy

It is understood that the consideration of some amount of SBG allows a photovoltaic system to uphold both its DB and the EB governing laws. The necessity of SBG, which can take any value between zero and one, raises the obvious question about its thermodynamically allowed value. To answer this question, we invoke the entropy balance (SB) of the photovoltaic system:

$$S_g + \Omega_s S_g^\infty(0, T_s) + \varepsilon_{SBG}\Omega_s S_0^g(0, T_s) = \Omega_c S_g^\infty(V, T) + \varepsilon_{SBG}\Omega_c S_0^g(0, T). \qquad (6)$$

Here, the non-negative irreversible entropy generation of the PV process is $S_g$. The entropy fluxes are given by [4,18]:

$$S_a^b(V,T) = \frac{2ke}{c^2h^3}\int_a^b E^2\left[\left(1+\frac{1}{exp\left(\frac{E-V}{kT}\right)-1}\right)ln\left(1+\frac{1}{exp\left(\frac{E-V}{kT}\right)-1}\right)\right. \tag{7}$$

$$\left.-\frac{1}{exp\left(\frac{E-V}{kT}\right)-1}ln\left(\frac{1}{exp\left(\frac{E-V}{kT}\right)-1}\right)\right]dE,$$

Minimizing the generated entropy $S_g$ from Eq. (6) allows us to calculate a unique $T$, $V$, and $\varepsilon_{SBG}$ that mutually solve the DB, EB, and SB laws. Treating Eqs. (1), (3), and (6) as the governing laws that determine the three free-variables $T$, $V$, and $\varepsilon_{SBG}$, we now arrive at a unified thermodynamic formulation of the photovoltaic effect.

## The unified model with examples

Let us move away from the limited case-study to consider a more general case of a work-producing cell with some heat conduction/convection to its environment. The DB law from Eq. (1) is only responsible for band-to-band processes, and therefore it is not explicitly affected by heat conduction $Q$. The consideration of $I$ and $Q$ enters the energy-balance law in the following way:

$$\Omega_s E_g^\infty(0,T_s) + \varepsilon_{SBG}\Omega_s E_0^g(0,T_s) = \Omega_c E_g^\infty(V,T) + \varepsilon_{SBG}\Omega_c E_0^g(0,T) + \tilde{E}I + Q, \tag{8}$$

The parameter $\tilde{E}$ denotes the average total energy (kinetic and potential) of the electron-hole pair. Likewise, the entropy balance is now modified as:

$$S_g + \Omega_s S_g^\infty(0,T_s) + \varepsilon_{SBG}\Omega_s S_0^g(0,T_s) = \Omega_c S_g^\infty(V,T) + \varepsilon_{SBG}\Omega_c S_0^g(0,T) + \tilde{S}I + \frac{Q}{T}, \tag{9}$$

With $\tilde{S}$ being the electron-hole average entropy. Equations (1), (8), and (9) allow us to tackle any situation where a photovoltaic system interacts with some environment. First, however, let us consider an isolated cell at an open circuit such that $Q = I = 0$, but now also with $\Omega_s = \Omega_c$. It is easy to see that a solution to this case emerged for any $\varepsilon_{SBG}$ once $T = T_S$, $V = S_g = 0$ as required for equilibrium conditions.

Next, let us consider a cell still at open circuit (*I=0*) but with heat conduction $Q$ to some environment. Taking a 300°K environment, the heat conduction term can be written as $Q = \sigma(T - 300)$ where $\sigma$ is the heat conduction coefficient. (Another source of heat removal could be the zero-potential emission (blackbody) from the back of the cell, which we will not treat here.) The temperature $T$, open circuit $V_{oc}$ and SBG $\varepsilon_{SBG}$ values of this case are plotted in Figure 2 as a function of the concentration factor $C = \Omega_s/\Omega_{s0}$ ($\Omega_{s0} = 6.85 \times 10^{-5}$ sr) and thermal conductivity $\sigma$ (W/m²K). It is seen that the temperature rises with the concentration factor and lowers with the conductivity value, as expected. Notice also that the SBG increases with temperature while the potential decreases. These two observations are in agreement with the known thermal behavior of photovoltaic devices[23]. Unlike past studies of the SBG[5–7], however, the sub-bandgap emissivity is now a state-variable of the system, like the temperature and the potential. As a result, the similar trend that $T$ and $\varepsilon_{SBG}$ experience is not explicit as it is usually assumed, as in Refs. 5–7, but emerges implicitly from the three governing laws of Eqs. (1), (8), and (9). Interestingly, Figure 2 shows that the cell cannot develop a potential unless it has some heat conduction to cool it down.

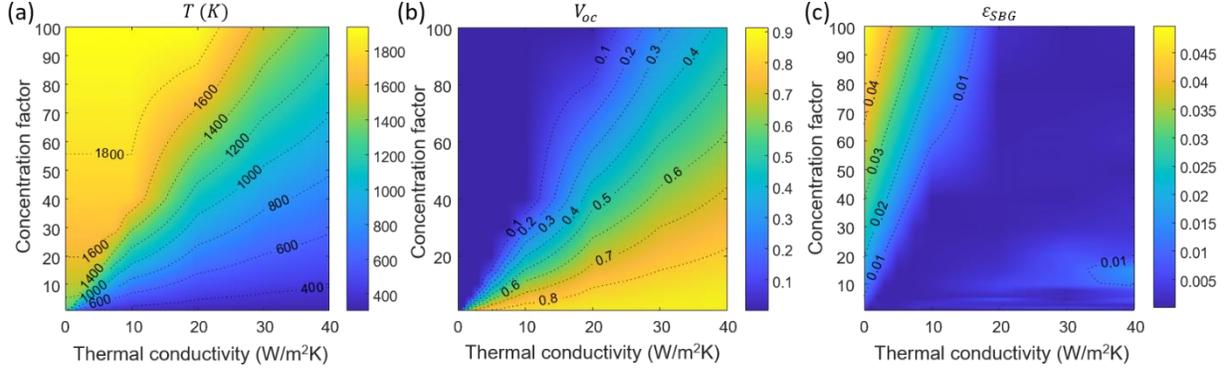

**Figure 2:** The temperature $T$ (a), open circuit potential $V_{oc}$ (b), and the sub-bandgap emissivity $\varepsilon_{SBG}$ (c), of a cell at open circuit as a function of the concentration factor and the thermal conductivity for $g = 1.12$ eV, and $T_S = 5778$ K.

Let us now investigate a cell doing work while conducting heat to its environment. To facilitate a solution to this case, we adopt the following straightforward definitions for the minority carriers' averaged energy and entropy:

$$\tilde{E} = \frac{E_g^\infty(V,T)}{N_g^\infty(V,T)} \; ; \quad \tilde{S} = \frac{S_g^\infty(V,T)}{N_g^\infty(V,T)} . \tag{10}$$

Figure 3(a) shows the cell's efficiency as a function of the bandgap and thermal conductivity. The efficiency is calculated as $\max(I \cdot V) / [\Omega_s \cdot E_0^\infty(0, T_s)]$. At high thermal conductivity, the efficiency approaches the SQ limit, the topmost row of Figure 3(a), which is ~30% for the blackbody source in this case. For lower values, however, the efficiency deviates significantly from the SQ prediction. The corresponding cell's temperature is shown in Figure 3(b). The temperature rises as the bandgap narrow, due to thermalization, and for a lesser heat-conduction. Therefore, according to our model, the maximal efficiency is a function of the cell's bandgap and heat conduction to its environment.

An interesting aspect of our model appearing in Fig. 3(b) is that the temperature begins to fall below a certain bandgap for a given thermal conductivity. This temperature drop happens since the maximum power point potential $V_m$ decreases as the bandgap narrow, as seen in Fig. 3(c). At some point $V_m = 0$, which is its minimal equilibrium value. Beyond this point, the cell must cool down to maintain DB. Figure 3(d) shows the SBG rises, in this case, to allow the system to uphold its EB constraint simultaneously. This character brings upon two possible operational regimes: the thermal one with zero potential and a finite amount of $\varepsilon_{SBG}$ and a PV one with a finite potential but with zero $\varepsilon_{SBG}$ (or at least one that approaches zero). These two regimes are separated by a state with $V_m \simeq \varepsilon_{SBG} \simeq 0$ in the bandgap-thermal conductivity plain. The white dotted line in Figure 3(a)-(d) depicts this 'zero-state' solution. The wavy nature of this line is an artifact of our computationally intensive numerical scheme. Figure 3(b) shows that this zero-state solution is also the 'maximal-temperature' solution. Most importantly, however, we have a thermodynamic prediction for the work available from a given PV process. In that sense, we go beyond past attempts to provide a thermodynamic justification for the detailed-balance law, attempts that were limited to open-circuit conditions ($I = 0$) where no work is possible[8–12].

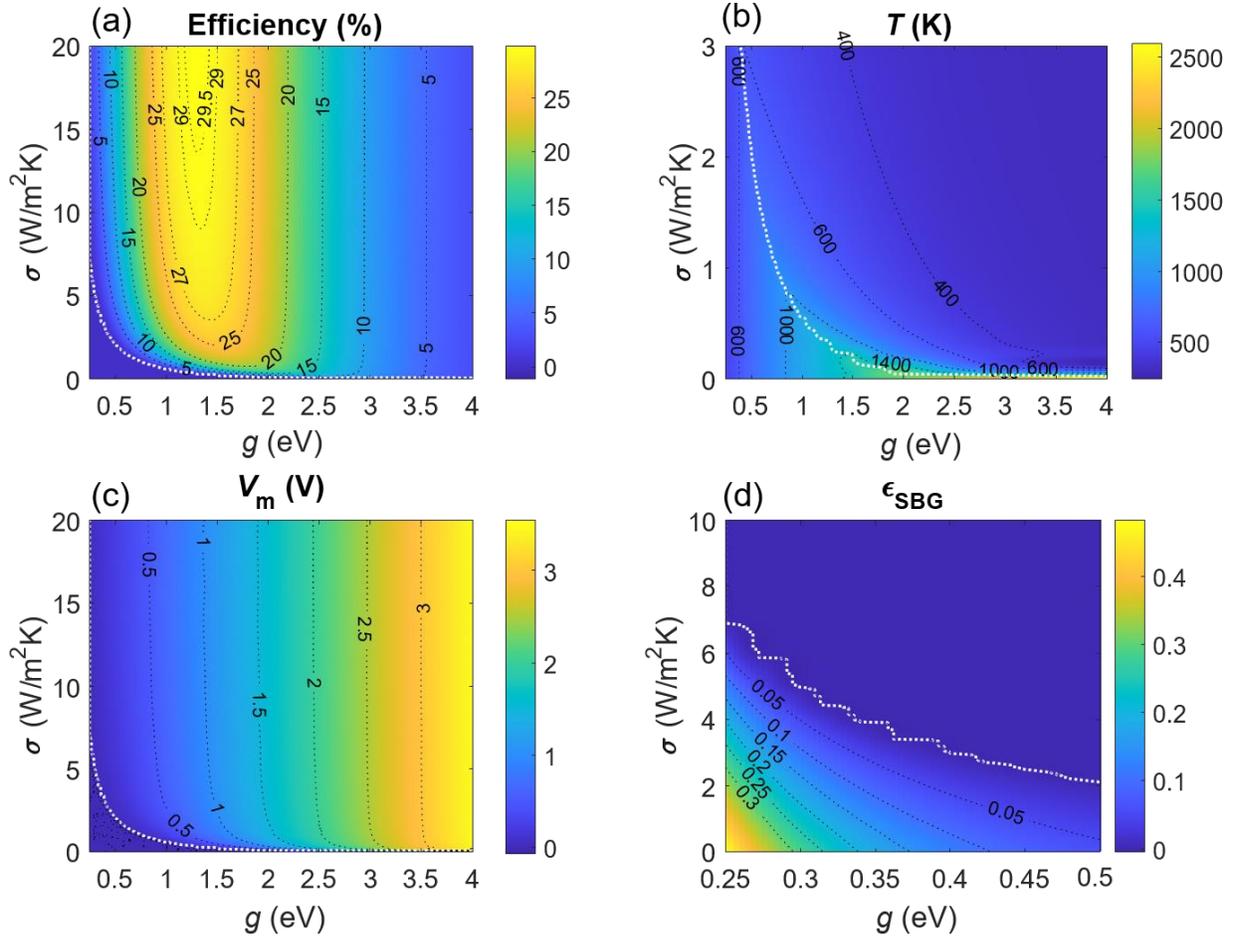

**Figure 3**: The efficiency, temperature $T$, $\varepsilon_{SBG}$, and maximal power point potential $V_m$ as a function of the bandgap $g$ and the thermal conductivity $\sigma$. The cell's parameters are $T_S = 5778$ K, $\Omega_S = 6.87 \times 10^{-5}$ sr and $\Omega_C = \pi$. The white dotted line shows the solution with $V_m = \varepsilon_{SBG} = 0$. Note that plots have different scales.

## Conclusions

We have shown that the thermodynamic state – temperature and potential – of an isolated extraterrestrial cell cannot always be determined at open circuit conditions using our present understanding of an idealized PV process. However, relaxing its ideality allowed us to determine its state at the expanse of having an additional free-variable in the form of the SBG. To specify this added free variable, we minimized the irreversible entropy production by the PV process. As a result, a unified description of the DB law with the first and second thermodynamics laws emerges. This model refines the SQ analysis and converges to it once the system's heat conduction is perfect. Our model identifies two operational regimes: the thermal one with finite SBG and vanishing potential and a PV one with a potential but with vanishing SBG. Among the two, only the second allows work from the cell. The transition between these two regimes is a function of the bandgap and the conductive/convective heat removal from the cell. Note that the SBG we find here is the minimal thermodynamic allowed value that any realistic SBG must exceed. This unified approach is particularly relevant to PV systems that are designated to operate at elevated temperatures, such as the thermophotovoltaics, thermoradiative, and thermophonic cells.

## Acknowledgments

The authors extend their gratitude toward Prof. Iris Visoly-fisher for her thoughtful comments regarding the manuscript's preparation.

## Funding

A. N. acknowledges the partial support of the Israel Science Foundation (ISF) no. 152/11 and the Adelis Foundation.

## Author Contributions

I. F. and A.N. developed the related theory, performed numerical calculations, and composed the manuscript. A.N. first recognized the inconsistency problem and its possible solution.

## Competing Interests

The authors declare no competing interest regarding the content of this article.

## Material & Correspondence

The custom code that supports this study's findings is available from the corresponding author upon reasonable request.